\documentstyle[aps,prl,epsf]{revtex}
\input epsf

\begin{document}

\title{Coordinates, modes and maps for the density functional}

\author{B.G. Giraud}
\address{giraud@dsm-mail.saclay.cea.fr, Service de Physique Th\'eorique, DSM,
CE Saclay, F-91191 Gif/Yvette, France}

\author{A. Weiguny}
\address{weiguny@uni-muenster.de, Institut f\"ur Theoretische Physik, 
Universit\"at M\"unster, Germany}

\author{L. Wilets}
\address{wilets@nuc2.phys.washington.edu, Box 351560, Department of 
Physics and Astronomy, University of Washington, Seattle, WA 98195-1560, USA}

\date{\today} 
\maketitle

\begin{abstract}

Special bases of orthogonal polynomials are defined, that are suited to 
expansions of density and potential perturbations under strict particle number 
conservation. Particle-hole expansions of the density response to an arbitrary 
perturbation by an external field can be inverted to generate a mapping 
between density and potential. Information is obtained for derivatives of 
the Hohenberg-Kohn functional in density space. A truncation of such 
an information in subspaces spanned by a few modes is possible. Numerical 
examples illustrate these algorithms.

\end{abstract}

\section{Introduction}

The well-known Hohenberg-Kohn (HK) density functional \cite{HK} and its finite
temperature generalization by Mermin \cite{Mer} suffer from the absence
of constructive algorithms after their respective existence theorems. The
Thomas-Fermi (TF) approach, however, and related developments such 
as \cite{BW}, \cite{KS}, have gone a long way into creating functionals with
practical physical values. For reviews on the effectiveness of the detailed
forms of the functional found empirically, see for instance \cite{DG} and
\cite{PK}. For applications of Skyrme forces to nuclear densities, see for 
instance \cite{JD} and \cite {GB}.

Standard perturbation theories (particle hole hierarchy of excitations, 
configuration mixing, generator coordinates, etc.), extrapolating from well 
understood mean field theories, give a constructive approach to the 
intricacies of a true ground state (GS), at the well known heavy
cost of calculations with many degrees of freedom. But such theories proved
to be practical, because suitable truncations were found that restricted 
calculations to few modes, collective or not, subspaces with fewer degrees 
of freedom. The purpose of the present note is to attempt answering a similar
question in the space of densities rather than the space of wave functions:
are there possible truncations, is there a possibility to restrict the 
functional to a set of few density modes?

For this we visit again the fundamentals of the HK functional $F[\rho]$ in 
a systematic approach, based upon the following chain of arguments, 

i) given the full Hamiltonian, $H=T+V+U,$ with a fixed kinetic operator 
$T=\sum_i t_i,$ a fixed two-body potential operator $V=\sum_{i>j} v_{ij},$ 
and a variable one-body potential operator $U=\sum_i u_i,$ assume a
non degenerate, square normalized GS $\Psi,$ with its corresponding 
eigenvalue $E,$ density $\rho$ and functional 
$F[\rho] \equiv \langle \Psi|(T+V)|\Psi\rangle = E -
\langle \Psi|U|\Psi\rangle;$ find the functional derivatives 
$\delta \rho/\delta u$ and, considering first $F$ as a functional of $u$ 
rather than $\rho,$ find $\delta F/\delta u,$ 

ii) expand such functional derivatives into suitable bases, to describe 
them by convenient matrices and vectors,

iii) then invert $\delta \rho/\delta u$ to know $\delta u/\delta \rho,$ 

iv) furthermore obtain $\delta F/\delta \rho$ by eliminating $\delta u$ 
between $\delta F/\delta u$ and $\delta u/\delta \rho;$ further information 
about $F$ might be obtained by integrating $\delta F/\delta \rho,$ or by 
comparing with  phenomenological approaches, such as gradient expansions,

v) at each stage, try a compression of the information,
by a truncation of the theory to a few ``density modes''.

A preliminary question is in order, however: can this formal program be 
carried if particle number is conserved in the mean only, as occurs with 
Lagrange multiplier techniques? According to \cite{DG}, the chemical 
potential, as a function of a continuous particle number, shows derivative 
discontinuities. We thus find it safer, in this paper, to stick to ``slices''
of the functional, those for fixed, integer particle numbers. We also restrict
our considerations to pure eigenstates of $H,$ at zero temperature.

In Section II, we carry the program in the absence of $V;$ the trivially 
soluble situation of independent fermions allows us to easily describe the 
mapping, $\rho \leftrightarrow u,$ for both infinitesimal and finite 
variations of $u.$ In Section III we reinstate $V,$ but use the Hartree-Fock
(HF) approximation to still obtain $\rho$ without excessive technical 
complications. Section IV is dedicated to a better understanding of
the ``tangent'' mapping $\delta \rho /\delta u$ \cite{LP} when full 
correlations are present. Then Section V introduces, via a new family of 
orthogonal polynomials, candidates for density and potential space modes, 
that might allow a compacted description of the functional and its 
$\rho \leftrightarrow u$ mapping. An investigation of the relevance of 
such modes is provided numerically, for a toy model of independent 
fermions. The numerical investigation is continued in Section VI, by means
of a second toy model, with now correlated fermions. Section VII contains
a discussion and a conclusion.

\section{Particle-hole expansions for independent fermions}

Let $Z$ be the particle number for a finite Fermion system. For simplicity 
we ignore discrete labels such as spins and isospins. Set $v=0,$ temporarily,
the case of two-body forces being discussed later. The one-body potential 
function $u$ is taken here as a local potential $u(r).$ 
The Hamiltonian then boils down to $H=\sum_{i=1}^Z (t_i+u_i)$ and its
GS, assumed to be non degenerate, is a Slater determinant. 
The HK functional, in this special case, reduces to the kinetic expectation 
value, $\langle \Psi | T | \Psi \rangle.$

Consider a perturbation $\delta u$ of the local potential $u.$ We find it 
practical to expand it in an orthonormal basis of functions $w_{\alpha}(r),$ 
namely 
$\delta u(r)=\sum_{\alpha=1}^{\infty} w_{\alpha}(r)\, \delta u_{\alpha}.$
According to the HK theorem \cite{HK}, this basis must be orthogonal to a 
``flat potential component'' $w_0(r)=1.$ This is satisfied if we find 
a basis such that, $\forall \alpha>0,\, \int dr\, w_{\alpha}(r)=0.$
It is then understood that the index $\alpha$ will run from $1$ to $\infty.$ 
For obvious practical reasons, however, the expansion will sooner or later 
be truncated at some finite rank of the basis.

The perturbation $\delta u$ induces a perturbation $\delta \rho$ of the 
density of the GS $\Psi$ of $H,$ and we find it convenient to 
expand $\delta \rho$ in the {\it same} orthonormal basis 
$\{w_{\beta}(r)\},$ namely 
$\delta \rho(r)=\sum_{\beta=1}^{\infty} w_{\beta}(r)\, \delta \rho_{\beta}.$
The fact that this basis satisfies the constraint, 
$\forall \beta>0,\ \int dr\, w_{\beta}(r)=0,$ is very useful
because $\delta \rho$ does not change the particle number, namely 
$\delta \rho$ automatically satisfies the condition 
$\int dr\, \delta \rho(r)=0.$ 
It is very convenient that the constraint of particle number conservation 
and that of ``non constant potential variation'' allow the same choice for 
our forthcoming basis $\{w_{\alpha}\}.$

A trivial particle-hole argument then provides that perturbation 
$\delta \Psi$ induced by $\delta u.$ Let the ``hole index'' $i=1, ..., Z$ 
and ``particle index'' $I$ (running from $Z+1$ to $\infty$) denote occupied 
and empty orbitals, respectively, with the corresponding single particle 
energies $\eta_i$ and $\eta_I$ and orthonormal wave functions 
$\psi_i$ and $\psi_I.$ For the sake of simplicity in the following, 
such orbital wave functions $\psi(r)$ are assumed to be real in the 
coordinate representation. Each filled orbital picks a variation 
$\delta \psi_i(r)=\sum_I \psi_I(r)\, 
\langle I|\delta u|i \rangle /(\eta_i-\eta_I).$
Hence,
\begin{equation}
\delta \rho(r)= 2 \sum_{iI} \, \psi_i(r) \psi_I(r) \,
\frac{\langle I | \delta u | i \rangle}{\eta_i-\eta_I} \, .
\label{prtrb1}
\end{equation}
This reads, when $\delta u$ and $\delta \rho$ are expanded,
\begin{equation}
\delta \rho_{\beta}= 2 \sum_{ iI\alpha} {\cal D}_{\beta iI} \,
\frac{ \langle I | w_{\alpha} | i \rangle } { \eta_i-\eta_I } \, 
\delta u_{\alpha} \, ,
\label{prtrb1xpd}
\end{equation}
where ${\cal D}$ denotes the projection of a particle-hole product of 
orbitals upon the basis $\{w_{\alpha}\},$
\begin{equation}
{\cal D}_{\beta iI} = \int dr \, w_{\beta}(r) \, \psi_i(r)\, \psi_I(r).
\label{connct}
\end{equation}
Note, incidentally, that particle-hole orthonormality ensures that, 
$\forall iI,$ $\int dr\, \psi_i(r) \psi_I(r)=0.$ Hence, functions 
$w_{\alpha}$ expanded in the basis of particle-hole products 
$\psi_i\, \psi_I,$ a basis to be orthonormalized, automatically fulfill the 
requested condition for $\delta u$ and $\delta \rho.$ Furthermore, positivity
of the density is guaranteed as variations $\delta \rho$ in Eq. (\ref{prtrb1})
are based on variations of the wave function, in particular by particle-hole 
admixtures to the GS determinant $\Psi.$

Define the matrix,
\begin{equation}
{\cal N}_{\beta \alpha} = 2 \sum_{iI} {\cal D}_{\beta iI} \, 
\frac{ \langle I | w_{\alpha} | i \rangle } { \eta_i-\eta_I } \, .
\label{defcurlN}
\end{equation}
Notice also that the perturbation matrix element, 
$\langle I | w_{\alpha} | i \rangle,$ coming from $\delta u,$
is nothing but an integral of a three term product, amounting to 
${\cal D}_{\alpha iI}.$ Notice finally that the energy denominators correspond
to a propagator $G=Q(E-QHQ)^{-1}Q,$ if Q is the particle-hole space 
projector. This operator is diagonal in the  particle-hole space, obviously. 
Then, in a condensed notation, ${\cal N}=2\, {\cal D}\, G\, \tilde {\cal D},$
where the tilde denotes transposition between index $\alpha$ and pair 
index $iI.$ Consider now $\delta \rho$ and $\delta u$ as just vectors with 
components $\delta \rho_{\beta}$ and $\delta u_{\alpha},$ respectively, 
sooner or later truncated. Since Eq. (\ref{prtrb1xpd}) reads 
$\delta \rho={\cal N} \delta u,$ the HK theorem
states that, under the usual condition of non degeneracy for the GS 
of $H,$ an inversion is possible. Namely for any $\delta \rho$ which 
leaves $\rho+\delta \rho$ in the manifold of {\it actual} densities, there 
exists a unique $u+\delta u,$ provided $\delta u$ does not add a constant 
component to $u.$ Under such precautions, the infinite matrix ${\cal N}$ can
be inverted, and the same can be expected under ``reasonable'' truncations of 
${\cal N}.$ Accordingly, while Eq. (\ref{prtrb1xpd}) provides the functional 
derivative $\delta \rho/\delta u,$ one obtains the functional derivative 
$\delta u/\delta \rho,$ 
\begin{equation}
\delta u_{\alpha}=\sum_{\beta} \left({\cal N}^{-1}\right)_{\alpha \beta} 
\delta \rho_{\beta}\, .
\label{invert}
\end{equation}

Now, that variation $\delta E$ of the GS energy induced by 
$\delta u$ is trivial. It just reads, 
\begin{equation}
\delta E=\langle \Psi | \delta U | \Psi \rangle,
\label{RaylRitz}
\end{equation}
because of the stationarity of the GS energy with respect to 
$\delta \Psi.$ Accordingly, for the functional under study,
\begin{equation}
\delta F = \delta E - \delta \langle \Psi | U | \Psi \rangle =
- \langle \delta \Psi | U | \Psi \rangle - 
  \langle \Psi | U | \delta \Psi \rangle = 
- 2 \langle \Psi | U | \delta \Psi \rangle =
- 2 \langle \Psi |U\, G\, \delta U | \Psi \rangle,
\label{short}
\end{equation}
hence,
\begin{equation}
\delta F = -2 \sum_{iI} \langle i | u | I \rangle\, (\eta_i-\eta_I)^{-1}\, 
\langle I | \delta u |i  \rangle,
\end{equation}
and finally, with proper expansions,
\begin{equation}
\delta F = -2 \sum_{iI} \langle i | u | I \rangle\, (\eta_i-\eta_I)^{-1}\, 
\sum_{\alpha} {\cal D}_{\alpha i I} 
\sum_{\beta} \left({\cal N}^{-1}\right)_{\alpha \beta} \delta \rho_{\beta}\, .
\label{prtrb2xpd}
\end{equation}

It may be convenient here to set from $u$ a column vector ${\cal U}$ with 
components
\begin{equation}
{\cal U}_{\, iI} = \int dr\, u(r)\, \psi_i(r)\, \psi_I(r),
\label{calU}
\end{equation}
hence the functional derivative $\delta F/\delta \rho$ reads, in a condensed, 
matrix and vector notation,
\begin{equation}
\delta F = -\, \tilde {\cal U}\, G\, \tilde {\cal D}\, \left( {\cal D}\, G\, 
\tilde {\cal D} \right)^{-1}\, \delta \rho.
\label{tool}
\end{equation}
This simplifies if one observes that, because of orthogonality 
between particle and hole orbitals, the product $ \psi_i(r)\, \psi_I(r)$ 
can be expanded in the $w$-basis as,
\begin{equation}
\psi_i(r)\, \psi_I(r) = \sum_{\beta} {\cal D}_{iI\beta}\, w_{\beta}(r).
\end{equation}
Accordingly,
\begin{equation}
{\cal U}_{iI}= 
\int dr\, u(r)\, \sum_{\beta} {\cal D}_{iI\beta}\, w_{\beta}(r) =
\sum_{\beta} u_{\beta}\, {\cal D}_{iI\, \beta},
\label{newrep}
\end{equation}
with $u_{\beta}$ the components of $u$ in our special basis. Combining 
Eqs. (\ref{tool}) and (\ref{newrep}) results in
\begin{equation}
\delta F = -\sum_{\alpha \beta \gamma} u_{\beta} 
\left( {\cal D}\, G\, \tilde {\cal D} \right)_{\beta \gamma}\, \left[ 
\left( {\cal D}\, G\, \tilde {\cal D} \right)^{-1} \right]_{\gamma \alpha}\,
\delta \rho_{\alpha} = -\sum_{\alpha} u_{\alpha}\, \delta \rho_{\alpha}.
\end{equation}
This avoids the transition between different bases through the matrix 
${\cal D}.$ One thus recovers the trivial result, 
$\delta F=-\int dr\, u(r)\, \delta \rho(r),$ but it must be kept in mind 
that, here, $u$ has become a functional of $\rho.$
Whether this simplification is made or not, this makes a set of numerical,
non linear, coupled, partial differential equations relating $F$ and $\rho.$ 
The non linearity comes in particular from the orbitals and single particle 
energies which occur in the definition of ${\cal N}.$ We stress again that 
the vector, $\tilde {\cal U}\, {\cal D}^{-1},$ just makes an ``$\alpha$'' 
representation of $u,$ converted from its particle-hole representation 
${\cal U}.$

A comment about Legendre transforms is here in order \cite{LP}. According to 
the Hellmann-Feynman theorem, $\delta E/\delta u=\rho.$ But then, 
$F \equiv E-\int dr\, u(r)\, \rho(r)$ is nothing but the Legendre transform 
of $E$ and the primary degree of freedom is not $u$ any more, but $\rho.$ 
Note that the reasoning remains if $V$ is reinstated. In all cases,
$u$ is recovered from $\delta F/\delta \rho=-u.$

For Eq. (\ref{tool}), and its generalization if two-body forces are 
present, to become a tool to obtain information about $F,$ dynamical
models are obviously necessary. These are the subjects of several of the 
forthcoming sections.

\section{Two-body forces and Hartree-Fock model}

In this section, we stay with $Z$ fermions, but reinstate in the Hamiltonian
the two-body interaction $v_{ij}$ with the operator 
$V=\sum_{i>j=1}^Z v_{ij}.$ The HK functional is 
$\langle \Psi | (T+V) | \Psi \rangle.$
While a Slater determinant $\Psi$ was available as 
the true GS of a simpler $H=T+U$ in the previous section, we cannot 
usually obtain the true GS with two-body forces present in a 
full $H=T+V+U.$ Thus, in this section, we tolerate for $\Psi$ the 
HF ground state of $H,$ with energy $E_0,$ and furthermore assume that this HF 
approximation does not create degeneracies between distinct $\Psi$'s. Under 
this precaution of uniqueness, there exists an extension of the HK theorem. 
Indeed, in the space of Slater determinants, let 
$\Psi$ and $\Psi'$ be the HF GSs of $H=T+V+U$ and $H'=T+V+U',$ 
respectively, and let $\rho(r)$ and $\rho'(r)$ be their respective densities. 
The two Hamiltonians differ by their (local) one-body operators 
$U=\sum_{i=1}^Z u(r_i)$ and $U'=\sum_{i=1}^Z u'(r_i)$ only. Their 
HF GS energies $E_0=\langle \Psi | H | \Psi \rangle$ and 
$E'_0=\langle \Psi' | H' | \Psi' \rangle,$ non degenerate, may 
be equal or distinct. Now, if $\rho$ and $\rho'$ were equal, then the 
usual HK arguments, namely 
$E_0'-E_0 < \langle \Psi  | (H'-H) | \Psi  \rangle =
\int dr\, [u'(r)-u(r)]\, \rho(r)$
and
$E_0-E_0' < \langle \Psi' | (H-H') | \Psi' \rangle =
\int dr\, [u(r)-u'(r)]\, \rho(r),$
necessarily lead to $\rho \ne \rho',$ by contradiction.

It can be stressed here that, again because of the stationarity of the energy 
with respect to variations of $\Psi,$ we can still take advantage of 
Eq. (\ref{RaylRitz}) for the variation of the energy induced by a variation 
$\delta u.$ This reads, with notations already used in the previous section,
\begin{equation}
\delta E_0= \langle \Psi | \delta U| \Psi \rangle\, .
\end{equation} 
The same holds every time we approximate the GS by means of 
the Rayleigh-Ritz principle in a restricted space of wave 
functions. As a general consequence, we obtain again Eq. (\ref{short}),
namely, $\delta F=-2\langle \Psi | U | \delta \Psi \rangle,$ for every such
variational approximation of $\Psi.$

That variation of $\delta \Psi$ induced by $\delta u$ is slightly 
more complicated, in the HF case, than in the trivial case of Section II 
where $H=T+U.$ 
Indeed, each filled orbital is driven by the perturbed HF equation,
\begin{eqnarray}
&& [\eta_i + \delta \eta_i - u(r) - \delta u(r)]\,
[\psi_i(r) + \delta \psi_i(r)] + \frac{\hbar^2}{2m}\, \Delta_r\,
[\psi_i(r) + \delta \psi_i(r)]\, = \nonumber \\
&& \sum_{j=1}^Z \int dr'\, v(r-r')\, [\psi_j(r') + \delta \psi_j(r')]^2\, 
[\psi_i(r) + \delta \psi_i(r)]\, - \nonumber \\
&& \sum_{j=1}^Z \int dr'\, v(r-r')\, [\psi_j(r') + \delta \psi_j(r')]\,
[\psi_i(r') + \delta \psi_i(r')]\, 
[\psi_j(r) + \delta \psi_j(r)]\, .
\label{prtbHF}
\end{eqnarray}
The non locality of the HF mean field, because of antisymmetrization, 
is written in an explicit way in the right-hand side above, in the coordinate
representation. An equivalent form of this perturbed HF Eq. (\ref{prtbHF}) 
is obtained if we retain its first order terms only and consider the 
particle-hole infinitesimal components $\delta c_{iI},$ again assumed here 
to be real numbers, 
\begin{equation}
(\eta_i-\eta_I)\, \delta c_{iI} - \langle I|\delta u|i \rangle =
\sum_{jJ}\, [\langle I J |v| i j \rangle + \langle I j |v| i J \rangle]\, 
\delta c_{jJ}.
\end{equation}
Thus $\psi_i(r)$ becomes $\psi_i(r)+\sum_I \psi_I(r)\, \delta c_{iI}.$ 
Notice that $\delta \eta_i$ drops out from the calculation, as 
should be expected. Then define in particle-hole space the symmetric matrix,
with antisymmetrized matrix elements of $v,$
\begin{equation}
A_{(iI)(jJ)}=(\eta_i-\eta_I) \delta_{ij} \delta_{IJ} -
\langle I J |v| i j \rangle - \langle I j |v| i J \rangle.
\end{equation}
Here $\delta$ is a Kronecker symbol and we must use pairwise indices
$(iI)$ when defining the inverse $A^{-1}$ to be used; this $A^{-1}$ 
generalizes the propagator used in Eq. (\ref{defcurlN}), 
and thus,
\begin{equation}
\delta c_{iI} = \sum_{jJ} \left(A^{-1}\right)_{(iI)(jJ)}\, 
\langle J|\delta u|j \rangle.
\end{equation}
This leads to the variation $\delta \rho,$ and the analog of 
Eq. (\ref{prtrb1xpd}) reads,
\begin{equation}
\delta \rho_{\beta}= 2 \sum_{(iI)(jJ)\alpha} {\cal D}_{\beta iI} \,
\left(A^{-1}\right)_{(iI)(jJ)}\, \langle J | w_{\alpha} | j \rangle \,
\delta u_{\alpha}, 
\end{equation}
where the overlap matrix ${\cal D}$ is the same as defined in the previous 
section. Actually, this boils down to the even simpler formula, in matrix
and vector notations,
\begin{equation}
\delta \rho = 2\, {\cal D}\, A^{-1}\, \tilde {\cal D}\, \delta u,
\end{equation}
where the tilde again denotes transposition of that connection ${\cal D}$
between the particle-hole products $\psi_i(r) \psi_I(r)$ and their 
rearrangement into an orthonormal basis $\{w_{\alpha}(r)\}.$ 
Note, incidentally, that, if $v=0,$ the matrix
${\cal F} \equiv 2\, {\cal D}\, A^{-1}\, \tilde {\cal D}$ boils down to 
the matrix ${\cal N},$ which is obviously negative semidefinite. We even 
expect ${\cal N}$ to be negative definite. The same is expected for 
${\cal D}\, A^{-1}\, \tilde {\cal D}.$ The stability of 
our HF solutions is assumed as long, at least, as $v$ is a weak 
enough interaction, and this ``definiteness'' of ${\cal F}$
is intuitively most likely. 

In the following, we shall also need the inverse of ${\cal F}.$ The final 
result for the variation of $F$ reads,
\begin{equation}
\delta F = -\, \tilde {\cal U}\, A^{-1}\, \tilde {\cal D}\, 
\left[\,{\cal D}\, A^{-1}\, \tilde {\cal D}\, \right]^{-1} \delta \rho,
\label{longfuncder}
\end{equation}
and, like in Section II, this expression, in a transparent notation, reads 
$\delta F=-\sum_{\alpha} u_{\alpha}\, \delta \rho_{\alpha}.$ For 
obvious reasons of numerical convergence, the number of needed 
$w_{\alpha}$ states must be large enough to overlap a sufficient number 
of particle-hole components of $\delta \rho.$ But as will be found 
in the coming numerical applications, a surprisingly small number 
of $w_{\alpha}$ states might sometimes be sufficient.

\section{Symmetry of the density-potential mapping in general}

With the exact ground energy $E$ and exact GS $\Psi$ of a full
$H=T+V+U,$ and $Q=1-| \Psi \rangle \langle \Psi |$ the projector
out of $\Psi,$ the Brillouin-Wigner perturbation theory gives the exact 
result for first order functional derivatives,
\begin{equation}
| \delta \Psi \rangle = \frac{Q}{E-QHQ}\, \delta U\, | \Psi \rangle =
\sum_{n=1}^{\infty} 
| \Psi_n \rangle\, \frac{\langle \Psi_n | \delta U | \Psi \rangle }{E-E_n}\,
.
\label{BW}
\end{equation}
We assume here that a resolution of the identity with real numbers and 
reasonable truncations, convenient for numerics, are available. The sum over 
excited states $\Psi_n$ includes integrals over the continuum, if necessary.
Let us single out the first of our identical particles and
integrate out all the other ones, to define the following transition densities,
\begin{equation}
\Theta_n(r)=\int dr_2\, dr_3\, ...\, dr_Z\, \Psi_n(r,r_2,r_3, ...,r_Z)\, 
\Psi(r,r_2,r_3, ...,r_Z).
\label{transdens}
\end{equation}
Notice that, from its very definition, $\Theta_n$ integrates out to $0,$ 
namely $\int dr\ \Theta_n(r)=0.$ Hence $\Theta_n$ can be represented in the 
$w$-basis without any loss of information.

Since $\delta U=\sum_{i=1}^Z \delta u(r_i)$ is a symmetric operator, it is 
clear that Eq. (\ref{BW}) also reads,
\begin{equation}
| \delta \Psi \rangle = \sum_{n=1}^{\infty} 
| \Psi_n \rangle\, \frac{Z \int dr\, \delta u(r)\, \Theta_n(r) }{E-E_n} =
\sum_{n\, \alpha}
| \Psi_n \rangle\, \frac{Z \int dr\, w_{\alpha}(r)\, \Theta_n(r) }{E-E_n}\,  
\delta u_{\alpha}\, ,
\label{BWr}
\end{equation}
where we have again expanded $\delta u$ in the basis $\{ w \}.$ There pops out
a matrix, 
\begin{equation}
D_{\alpha\, n}=Z \int dr\, w_{\alpha}(r)\, \Theta_n(r),
\label{gencurlD}
\end{equation}
as a generalization of the matrix ${\cal D}_{\alpha\, iI}.$

Now, by definition, the density of the GS is,
\begin{equation}
\rho(r)= 
Z \int dr_2\, dr_3\, ...\, dr_Z\, \left[\Psi(r,r_2,r_3, ...,r_Z)\right]^2,
\end{equation}
and its variation is,
\begin{equation}
\delta \rho(r)= 
2 Z \int dr_2\, dr_3\, ...\, dr_Z\, \Psi(r,r_2,r_3, ...,r_Z)\, 
                             \delta \Psi(r,r_2,r_3, ...,r_Z).
\end{equation}
This becomes, if one replaces $\delta \Psi$ by its expression, Eq. (\ref{BWr}),
\begin{equation}
\delta \rho(r)= 
2 Z \sum_{n\, \alpha} \Theta_n(r)\, \frac{D_{\alpha\, n}}{E-E_n}\,  
\delta u_{\alpha}.
\end{equation}
An expansion of $\delta \rho$ in the $\{ w \}$ basis gives its coordinates,
\begin{equation}
\delta \rho_{\beta}= 
2\, \sum_{n\, \alpha} D_{\beta\, n}\, \frac{1}{E-E_n}\, D_{\alpha\, n}\, 
\delta u_{\alpha},
\label{genflex}
\end{equation}
hence, in an obvious notation, a {\it symmetric} ``flexibility'' matrix
${\cal F}=2\, D\, G\, \tilde D$ connecting $\delta u$ and $\delta \rho.$ In 
hindsight, the symmetry of ${\cal F}$ (and of its approximations under the 
Rayleigh-Ritz variational principle) is straightforward. Indeed, since 
$u=-\, \delta F/\delta \rho,$ then 
$\delta u_{\alpha} / \delta \rho_{\beta} = 
-\, \delta^2 F / (\delta \rho_{\alpha}\, \delta \rho_{\beta}).$ 
All denominators $E-E_n$ being negative definite, the negative definite nature
of this exact ${\cal F}$ is also transparent.

We conclude this section on the general case with explicit expressions for 
$\delta F/\delta \rho$ and $\delta^2 F/(\delta \rho \delta \rho').$ With 
Eqs. (\ref{BWr}), (\ref{gencurlD}) the general form for $\delta F,$ see 
Eq. (\ref{short}), reads
\begin{equation}
\delta F = -2 \langle \Psi | U | \delta \Psi \rangle = -2 \sum_{n \alpha}
\langle \Psi | U | \Psi_n \rangle\, (E-E_n)^{-1} D_{\alpha n}\, 
\delta u_{\alpha} = -2 \sum_{n \alpha} {\cal U}_n\, (E-E_n)^{-1}
 D_{\alpha n}\, \delta u_{\alpha}.
\end{equation}
Here the numbers
\begin{equation}
{\cal U}_n = 
\int dr\, \langle \Psi | \left[ \sum_{i=1}^Z u(r_i)\right] | \Psi_n \rangle =
Z \int dr\, u(r)\, \Theta_n(r),
\end{equation}
are now the components of $u$ in the space of transition densities, 
generalizing Eq. (\ref{calU}) for states $\Psi,$ $\Psi_n$ containing 
correlations. Upon inverting Eq. (\ref{genflex}) we find, as a generalization 
of Eq. (\ref{longfuncder}),
\begin{equation}
\delta F = -\, \tilde {\cal U}\, G\, \tilde D\, 
\left[\, D\, G\, \tilde D\, \right]^{-1} \delta \rho.
\label{geneof19}
\end{equation}
This simplifies if we expand
\begin{equation}
\Theta_n(r) = Z^{-1} \sum_{\beta}  D_{n \beta}\, w_{\beta}(r).
\end{equation}
Then
\begin{equation}
{\cal U}_n = \int dr\, u(r) \sum_{\beta} D_{n \beta}\, w_{\beta}(r) = 
\sum_{\beta} u_{\beta}\, D_{n \beta},
\end{equation}
with again the components of $u$ in our special $w$-basis, 
$u_{\beta}=\int dr\, u(r)\, w_{\beta}(r).$ Accordingly,
\begin{equation}
\left[ \tilde {\cal U}\, G\, \tilde D \right]_{\gamma} = \sum_{\beta n}
u_{\beta}\, D_{\beta n}\, G_{nn}\, \tilde D_{n \gamma} = \frac{1}{2}
\sum_{\beta} u_{\beta}\, {\cal F}_{\beta \gamma},
\end{equation}
hence finally, as expected,
\begin{equation}
\delta F = - \sum_{\alpha \beta \gamma} u_{\beta}\, {\cal F}_{\beta \gamma}\,
\left[ {\cal F}^{-1} \right]_{\gamma \alpha}\, \delta \rho_{\alpha} =
- \sum_{\alpha} u_{\alpha}\, \delta \rho_{\alpha},
\end{equation}
as before in sections II and III. Similarly, the second derivative of $F$ is 
found directly from the inverse of Eq. (\ref{genflex}),
\begin{equation}
\delta^2 F/(\delta \rho_{\alpha} \delta \rho_{\beta}) =
- \left( {\cal F}^{-1} \right)_{\alpha \beta} = 
- \left[ \left( 2 D G \tilde D \right)^{-1} \right]_{\alpha \beta}.
\label{scndrv}
\end{equation}
Hence, from Eqs. (\ref{geneof19}) and (\ref{scndrv}),
\begin{equation}
\delta F/\delta \rho_{\alpha} = 2 \sum_{\beta} 
\left( \tilde {\cal U} G \tilde D \right)_{\beta}\ 
\delta^2 F / (\delta \rho_{\beta} \delta \rho_{\alpha}),
\end{equation}
another useful equation to test phenomenological functionals $F[\rho],$ 
by calculating quantities such as $\tilde {\cal U} G \tilde D$ from 
microscopic wave functions and energies for simple systems.

\section{One dimensional toy model, special polynomials}

Assume that $r$ is just one dimensional, running from $-\infty$ to $\infty.$ 
Define $t=-d^2/(2dr^2),$ with a nucleon mass $m=1,$ $\hbar=1$ and $p=-id/dr.$
In the present section, we are first interested in the Hamiltonian 
$H_0=\sum_{i=1}^Z (p_i^2/2+ r_i^2/2);$ it is not a bad approximation to
most shell model Hamiltonians, whether one considers one-body potentials only 
or HF solutions to problems with two-body potentials as well. A 
trivial scaling of coordinates and momenta allows us to reduce to the case, 
$\omega=1,$ any situation, $\sum_i^Z (p_i^2/2+\omega^2  r_i^2/2),$ where the 
physical spring constant would be different. 

Then we shall consider the functional 
$F(\rho)=\langle \Psi | H_0 | \Psi \rangle$ for a family of additional 
one-body potentials $u,$ with the corresponding GS density $\rho(r)$ of 
\begin{equation}
H=\sum_{i=1}^Z\, [\, t_i+r_i^2/2+u(r_i)\, ].
\end{equation}
Set temporarily $u=0,$ namely consider $H_0$ and its GS density 
$\rho_0(r).$ Since $\omega=1,$ which will be understood from now on, both 
initial particle and hole orbitals $\psi_k(r)$ are just trivial products 
$\varphi_k$ of a Hermite polynomial, a common Gaussian and a suitable 
normalization, 
$\varphi_k(r)=\pi^{-\frac{1}{4}}\, e^{-\frac{1}{2} r^2}\, P_k(r).$ 
For the sake of illustration, we list here the first five Hermite 
polynomials, with their coefficients adjusted for orthonormalization,
\begin{equation}
P_0=                       1,  \, 
P_1= 2^{\frac{1}{2}}\,     r,  \,  
P_2=
   2^{-\frac{1}{2}}\,   (2r^2-1), \, 
P_3=
  3^{-\frac{1}{2}}\, r\,(2r^2-3), \, 
P_4=2^{-1}\, 6^{-\frac{1}{2}}\,(4r^4-12r^2+3).
\label{Herm}
\end{equation}
Particle-hole products, $\varphi_i(r)\, \varphi_I(r),$ make, in turn, 
just polynomials again, now multiplied by $e^{-r^2}.$ To build our 
basis, $\{w_{\alpha}(r)\},$ it is tempting to orthonormalize the set 
$\{\varphi_i \varphi_I\}$ containing that Gaussian, $e^{-r^2},$ and recover 
forms $2^{\frac{1}{4}}\varphi_k(r \sqrt2),$ with Hermite polynomials again, 
compressed by the obvious $\sqrt 2$ for their argument $r,$ because of the 
new factor, $e^{-r^2}.$ This is correct for odd parity functions. But, for 
even parity ones, the constraint, $\int dr\, w_{\alpha}(r)=0,$ would be 
violated. Hence, out of each even function, 
$2^{\frac{1}{4}}\varphi_{2k}(r \sqrt2),\, k>0,$ 
we subtract a term proportional to $2^{\frac{1}{4}}\varphi_0(r \sqrt2),$ 
letting the subtraction cancel the integral, $\int_{-\infty}^{\infty}.$ 
(Alternately, we considered all elementary functions $r^{2k}\, e^{-r^2}.$) 
Then we use a Gram-Schmidt process to reorthonormalize such 
subtracted states. Notice that the subtraction cancels out the polynomial 
state of degree zero, and therefore the transformation from Hermite 
polynomials to this new set of orthonormal polynomials is not unitary, but 
only isometric, with ``defect index'' $1.$ In other words, our basis has 
codimension $1.$ This is also clear from the degree $2$ of the lowest member 
of the new even basis. For an illustration, we list the first four even 
states obtained,
\begin{eqnarray}
w_2(r)=
2\ 2^{\frac{1}{4}}\,( 2\,r^2-1)/ \sqrt3 \ 
\pi^{-\frac{1}{4}}\, e^{-r^2}, \nonumber \\
w_4(r)= 
2^{\frac{3}{4}}\,( 8\,r^4-14\,r^2+1)/ \sqrt{15} \ 
\pi^{-\frac{1}{4}}\, e^{-r^2}, \nonumber \\
w_6(r)= 
\left( 32\,r^6 - 128\,r^4 + 94\,r^2  -11 \right)/
\left(2^{\frac{1}{4}} \sqrt{105}\right) \ 
\pi^{-\frac{1}{4}}\, e^{-r^2}, \nonumber \\
w_8(r)= 
\left( 128\,r^8 - 928\,r^6 + 1752\,r^4 - 906\,r^2 + 39\right)/
\left(9\ 2^{\frac{3}{4}} \sqrt{35}\right)\ 
\pi^{-\frac{1}{4}}\, e^{-r^2}.
\label{weightr2}
\end{eqnarray}
For the sake of comparison with Hermite polynomials, which rather go 
with a factor $e^{-\frac{1}{2}r^2},$ we perform the 
transformation, $r \rightarrow r/\sqrt{2}$ on Eqs. (\ref{weightr2})
and multiply the results by a factor $2^{-\frac{1}{4}}$ to retain 
their (ortho)normalization. Discarding norm coefficients from the resulting 
polynomials we get,
\begin{eqnarray}
Q_2=r^2-1, \nonumber \\
Q_4=2\,r^4-7\,r^2+1, \nonumber \\
Q_6=4\,r^6-32\,r^4+47\,r^2-11, \nonumber \\
Q_8=8\,r^8-116\,r^6+438\,r^4-453\,r^2+39.
\label{weightr2/2}
\end{eqnarray}
But, as already noticed, products $\varphi_i\, \varphi_I$ carry a factor 
$e^{-r^2}$ and we find it natural, in the following, to stick to those 
polynomials trivially derived from Eqs. (\ref{weightr2}),
\begin{eqnarray}
\Gamma_2(r)=
2\, (2\,r^2-1)/ \sqrt3\, , \nonumber \\
\Gamma_4(r)= 
2^{\frac{1}{2}}\, (8\, r^4 - 14\, r^2 + 1)/ \sqrt{15}\, , \nonumber \\
\Gamma_6(r)= 
\left(32\, r^6 - 128\, r^4 + 94\, r^2  -11 \right)/ \sqrt{210}\, , \nonumber 
\\
\Gamma_8(r)= 
\left( 128\,r^8 - 928\,r^6 + 1752\,r^4 - 906\,r^2 + 39\right)/
\left(18\, \sqrt{35}\right)\, ,
\label{polevn}
\end{eqnarray}
and so on. We generated such polynomials up to degree 100 and will send them 
to interested readers. Such polynomials are orthonormal under the metric 
weight, $e^{-2r^2}\, \sqrt{2/\pi}.$ They must be completed by odd 
Hermite polynomials, $P_{2k+1}(r\sqrt2),$  suitably adjusted for the same 
metric. Hence, for instance,
\begin{mathletters}
\begin{eqnarray}
\Gamma_1(r)=2\, r\, , \\
\Gamma_3(r)= r\, (4r^2-3)\, \sqrt{2/3}\, , \\
\Gamma_5(r)= r\, (16r^4-40r^2+15)/ \sqrt{30}\, , \\
\Gamma_7(r)= r\, (64r^6-336r^4+420r^2-105)/(6\, \sqrt{35})\, .
\label{polevod}
\end{eqnarray}
\end{mathletters}
More technicalities on such polynomials $\Gamma$ and related polynomials
can be found in \cite{GiMeWe}.

It is then trivial to calculate both even and odd blocks, respectively, 
of the initial matrix ${\cal D},$ see Eq. (\ref{connct}), according to the 
parity of the subscript of $w$ and its associated polynomial $\Gamma.$ With 
due normalizations, this reads,
\begin{equation}
{\cal D}_{\beta i I}=\int_{-\infty}^{\infty} dr\ 
\left[ \Gamma_{\beta}(r)\, e^{-r^2}\,
(2/\pi)^{\frac{1}{4}} \right]\ 
\left[ P_i(r)\, e^{-\frac{1}{2}r^2} \pi^{-\frac{1}{4}} \right]\ 
\left[ P_I(r)\, e^{-\frac{1}{2}r^2} \pi^{-\frac{1}{4}} \right]\, .
\end{equation}
For instance, if the hole label is 
restricted to $i=0,$ and the particle label $I$ runs from $1$ to 
$3,$ the sets of non vanishing odd, respectively even, matrix elements boil 
down to,
\begin{equation}
{\cal D}_{101}= 2^{-\frac{3}{4}}\, \pi^{-\frac{1}{4}},\, 
{\cal D}_{103}=-\sqrt{3}/\left[ 4\, (2 \pi)^{1/4} \right],\,  
{\cal D}_{303}= 2^{-\frac{7}{4}}\, \pi^{-\frac{1}{4}},\ \ \ \ \ 
{\cal D}_{202}=\sqrt{3}\ 2^{-\frac{7}{4}}\, \pi^{-\frac{1}{4}}.
\end{equation}

We found it useful to precalculate and store such initial matrix 
elements ${\cal D}$ for the particle index $I$ running up to 100 and the 
$\alpha$ index running up to 100 also. This fastens generic calculations
of ${\cal D}$ when $u$ becomes finite, as one represents $(t+r^2/2+u)$ by a 
matrix on the oscillator basis, diagonalizes it with eigenvalues 
$\eta_k$ and orthonormal eigenvectors $X_{\ell k},$ and finally 
expands orbitals of both holes and particles as 
$\psi_k(r)=\sum_{\ell} X_{\ell k}\, \varphi_{\ell}(r)$ in the same basis.

In \cite{GiMeWe} we set $Z=4,$ considered $u$ to be an infinitesimal 
$\delta u$ in the neighborhood of $u=0,$ then calculated and diagonalized the 
functional derivative ${\cal N}=\delta \rho/\delta u.$ The eigenvectors of 
${\cal N}$ defined density and potential infinitesimal perturbations having 
the same shapes. Now we set again $Z=4,$ but are rather interested in 
cases where $u$ is finite. We are concerned in particular with the mapping 
between $u$ and $\rho,$ in that representation provided by the ``modes'' 
$w_{\alpha}.$ Truncations at a maximum degree $N$ are necessary. The finite 
expansion,
\begin{equation}
u(r) = \sum_{\alpha=1}^N u_{\alpha}\, w_{\alpha}(r),
\end{equation}
defines those processed perturbations $u.$ Given $u,$ it is 
trivial to diagonalize $H$ with a good numerical accuracy and obtain $\rho.$ 
Then it is easy to obtain ``coordinates in density space,''
\begin{equation}
\rho_{\alpha} = \int_{-\infty}^{\infty} dr\, w_{\alpha}(r)\, 
[\rho(r)-\rho_0(r)].
\end{equation}
The harmonic potential, $r^2/2,$ serves here as the origin in 
potential space, and the origin in density space is the corresponding 
density $\rho_0.$ We show in Figures 1 and 2, respectively, a grid of values
$\{u_2,u_4\}$ in potential space and its image grid of density coordinates 
$\{\rho_2,\rho_4\}.$ Dots at grid corners help matching the object and the 
image.

\begin{figure}[htb] \centering
\mbox{  \epsfysize=100mm
         \epsffile{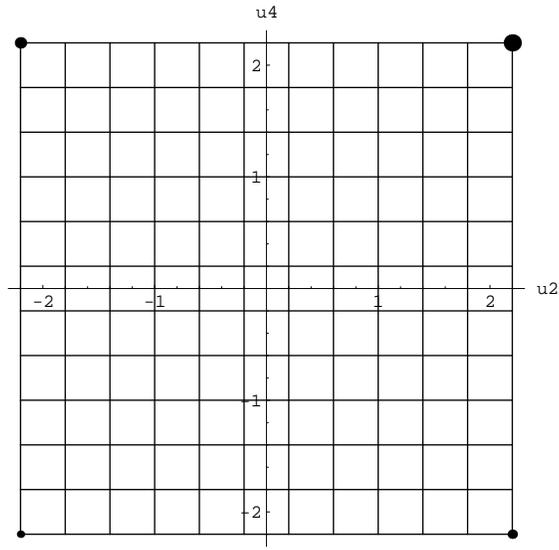}
     }
\caption{Grid of parameters $u_2,u_4$ for the potential 
$u=u_2\, w_2+u_4\, w_4$ used in the toy model.}
\end{figure}

\begin{figure}[htb] \centering
\mbox{  \epsfysize=100mm
         \epsffile{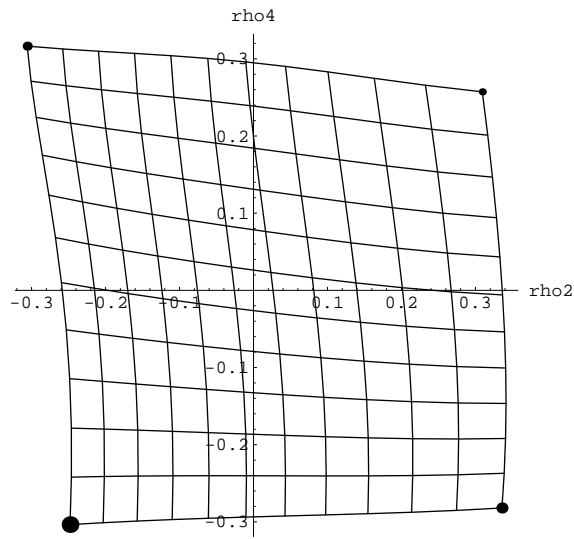}
     }
\caption{Density space image, projected onto the $\rho_2,\rho_4$ plane, of 
the grid of potentials of Fig. 1.}
\end{figure}

\begin{figure}[htb] \centering
\mbox{  \epsfysize=100mm
         \epsffile{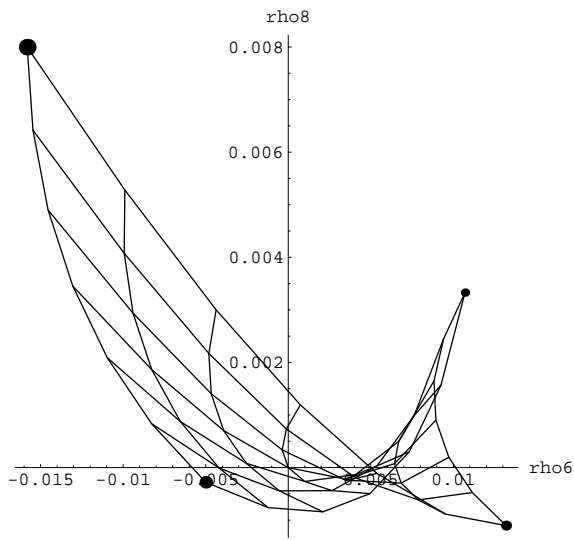}
     }
\caption{Same as Fig. 2, but now the image grid is projected onto the 
$\rho_6,\rho_8$ plane.}
\end{figure}

\begin{figure}[htb] \centering
\mbox{  \epsfysize=100mm
         \epsffile{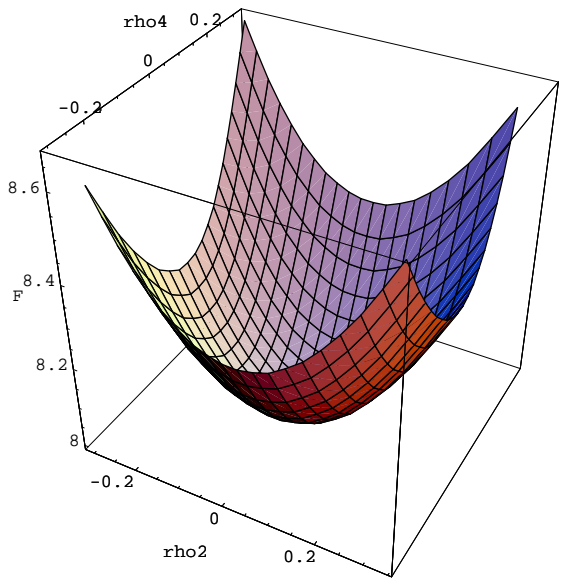}
     }
\caption{Toy model HK functional in a $\{\rho_2,\rho_4\}$ frame. Note small 
deviations from paraboloid.}
\end{figure}

In this calculation, all coordinates $u_{\alpha}$ have been set to vanish, 
except $u_2$ and $u_4,$ but it must be stressed that the resulting density 
variation, $\rho-\rho_0,$ has non vanishing coordinates $\rho_6,\rho_8,...$ 
besides $\rho_2$ and $\rho_4.$ Such additional coordinates are small, but not
very small, as shown by the grid for $\rho_6,\rho_8$ in Figure 3.
Qualitatively, if $u$ contains one mode $w_{\alpha}$ only,
then $\rho_{\beta}$ tends to decrease when $|\alpha-\beta|$ increases. 
But this is likely to be valid for small enough $u$'s only, in a linear 
response regime. Curvature effects, evidenced by Figs. 2 and 3, 
must be expected further.

Of interest are plots of $F$ in $\rho$-space. If $u$ has two 
components $u_2,u_4$ only, assume that $F$ is a function of $\rho_2,\rho_4$ 
only. Then a gradient $\nabla F(\rho)$ can be observed directly. In Figure 4, 
the 3D plot of $F$ shows slight deviations from a traditional paraboloid. 
This is even more visible in Figure 5, showing the vector field 
$\{u_2,u_4\}(\rho_2,\rho_4),$ namely $-\nabla F.$ The field, read from 
Figs. 2 and 1, focuses towards the origin in $\rho$-space, but with 
clear distortions. We know that the field has a vanishing curl; it 
can be integrated back into $F.$

It is also trivial to create an approximate $F$ in the following way: i) 
assume indeed that $F$ depends only on $\rho_2$ and $\rho_4$ for this toy 
model, ii) take a few exact (numerical, actually) values of $F$ at random 
points taken from the partner grids shown in Figs. 1 and 2, iii) set a 
simple parametric ansatz such as,
\begin{eqnarray}
F_{app} \simeq F_{00} + F_{10}\, \rho_2 &+& F_{01}\, \rho_4 + \left( F_{20}\, 
\rho_2^2 + 2\, F_{11}\, \rho_2\, \rho_4 + F_{02}\, \rho_4^2 \right)/2 + 
\left( F_{30}\, \rho_2^3 + 3\, F_{21}\, \rho_2^2\, \rho_4 + 3\, F_{12}\, 
\rho_2\, \rho_4^2 + F_{03}\, \rho_4^3 \right)/6 + 
\nonumber \\
&& \left(F_{40}\, \rho_2^4\, + 4\, F_{31}\, \rho_2^3 \rho_4 + 6\, F_{22}\, 
\rho_2^2\, \rho_4^2 + F_{13}\, \rho_2\, \rho_4^3 + F_{04}\, \rho_4^4 \right)
/24 \, ,
\label{Fapp}
\end{eqnarray}
and, finally, iv) least square fit the ``exact'' values selected at step 
ii). There are here 15 parameters and it is reasonable to select typically 
about twice as many exact values for the least square fit. The following 
result,
\begin{eqnarray}
F_{app} \simeq 8.0005176 - .0026263\, \rho_2 + 3.7994711\, \rho_2^2 - 
.9776987\, \rho_2^3 - .0208120\, \rho_2^4 + .0034190\, \rho_4 + 
\nonumber \\
.9487531\, \rho_2\, \rho_4 - .2536905\, \rho_2^2\, \rho_4 - 
.5398019\, \rho_2^3\, \rho_4 + 3.7854603\, \rho_4^2 + 
.9596720\, \rho_2\, \rho_4^2 + 
\nonumber \\
 .0368183\, \rho_2^2\, \rho_4^2 + .0172911\, \rho_4^3 - 
 3.0193024\, \rho_2\, \rho_4^3 - .2356997\, \rho_4^4\, ,
\label{numFapp}
\end{eqnarray}
comes from fitting 26 values for $Z=4.$ Figure 6 shows several resulting 
contours, the smallest of which locates the minimum of $F_{app}$ very slightly 
only away from the origin, that is the true minimum by the very construction 
of the toy model. The value of the functional at the minimum turns out to be 
$8.0005165,$ instead of strictly $8.$ This toy numerical exercise 
demonstrates the possibility of contracting the description of the functional 
to few degrees of freedom.

\begin{figure}[htb] \centering
\mbox{  \epsfysize=100mm
         \epsffile{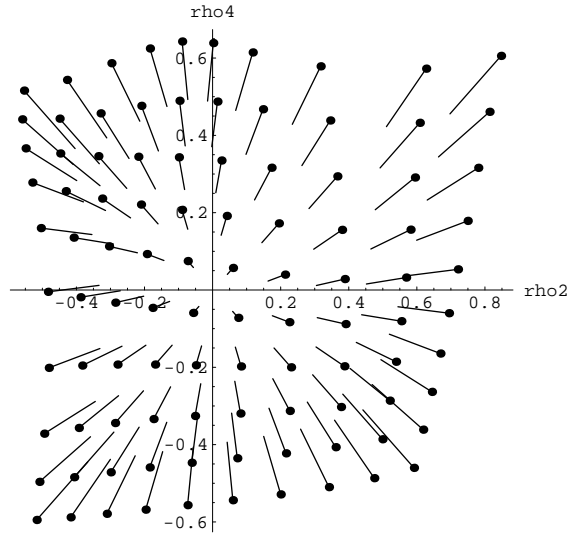}
     }
\caption{Same as Fig. 4. The lines represent $-\nabla F$ at points  
$\{\rho_2,\rho_4\}$ shown by dots. Note deviations from radial pattern.}
\end{figure}

\begin{figure}[htb] \centering
\mbox{  \epsfysize=100mm
         \epsffile{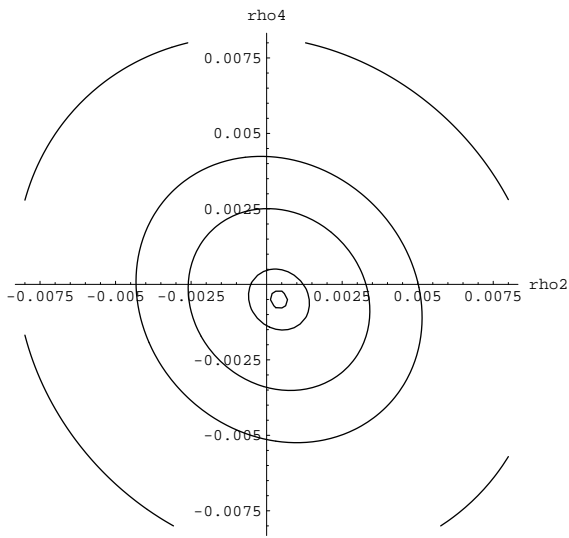}
     }
\caption{Contours for $F_{app}.$}
\end{figure}

\section{Correlations, from another toy model}

In the previous sections we skirted around the difficulty of obtaining a 
GS with true correlations. Now we shall mix several Slater determinants, 
each made of $Z=4$ harmonic oscillator, one dimensional orbitals taken from 
$h_0=\frac{1}{2}(p^2+r^2).$ The Hamiltonian is a complete one, 
\begin{equation}
H=\sum_{i=1}^4 \left[\frac{1}{2} \left(-\frac{d^2}{dr_i^2}+r_i^2\right) +
u_2\, w_2(r_i)+u_4\, w_4(r_i)\right]  - V_a \sum_{i>j=1}^4 
\left[\delta(r_i-r_j-R_a)+\delta(r_i-r_j+R_a)\right] .
\end{equation}
This contact, finite range attraction between particles is expected to create 
a reasonable amount of correlations and was chosen to allow an easy 
precalculation and tabulation of matrix elements of $v.$ Such matrix elements 
are again understood to be antisymmetrized.

The first Slater determinant, $\Phi_0,$ in the mixture contains the lowest $Z$ 
orbitals of the harmonic oscillator. It is expected to make the dominant 
component of the configuration mixture $\Psi,$ as we shall keep $u_2,$ $u_4$ 
within the grid seen in Fig. 1, and also the strength $V_a$ moderate. Let 
$\xi_i$ and $\varphi_j,$ $i,j=1,...,Z$ be the orbitals of two Slater 
determinants $\Xi$ and $\Phi,$ respectively. Define the cofactors $C_{ij}$ and 
double cofactors $C_{ikjl}$ of the determinant of scalar products 
$\langle \xi_i | \varphi_j \rangle.$
Such cofactors are very simple in the present orthogonal basis of orbitals, 
obviously. Then the matrix elements needed for the Hamiltonian matrix and the 
calculation of the HK functional read,
\begin{equation}
\langle \Xi | (H_0+U) | \Phi \rangle =
\sum_{ik} C_{ik}\, \langle \xi_i | (h_0+u) | \varphi_k \rangle,\ \ \
\langle \Xi | V | \Phi \rangle = \frac{1}{4} \sum_{ijkl}  C_{ikjl}\, 
\langle \xi_i \xi_j | v | \varphi_k \varphi_l \rangle
\, .
\end{equation}

With $V_a=5$ and $R_a=1$ the grids shown by Figures 7 and 8 come from a 
calculation with a single particle basis made of the first 10 harmonic 
oscillator orbitals and a corresponding Slater basis of 61 states made 
of $\Phi_0$ and all positive parity one-particle-one-hole and 
two-particle-two-hole determinants built upon $\Phi_0.$ Because of $V$ the 
centers of the density grids are not at the origin defined by $\Phi_0,$ 
obviously. Indeed, for $u_2=u_4=0,$ this calculation gives 
$\{\rho_2,\rho_4,...\rho_{14}\}=\{-.14,-.36,-.43,-.06,.04,-.01,.005\}$ as
the coordinates of the ground state density shift $\rho-\rho_0$ and the 
ground state energy is $E=-6.7,$ significantly down from 
$\langle \Phi_0 | H_0 | \Phi_0 \rangle =8.$

\begin{figure}[htb] \centering
\mbox{  \epsfysize=100mm
         \epsffile{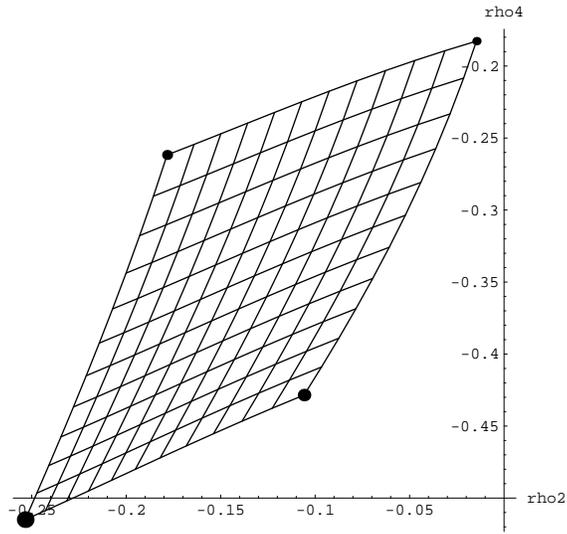}
     }
\caption{Second toy model: influence of the two-body force $V$ on 
the $\rho_2,\rho_4$ image of the grid of Fig. 1. Compare with Fig. 2.}
\end{figure}

\begin{figure}[htb] \centering
\mbox{  \epsfysize=100mm
         \epsffile{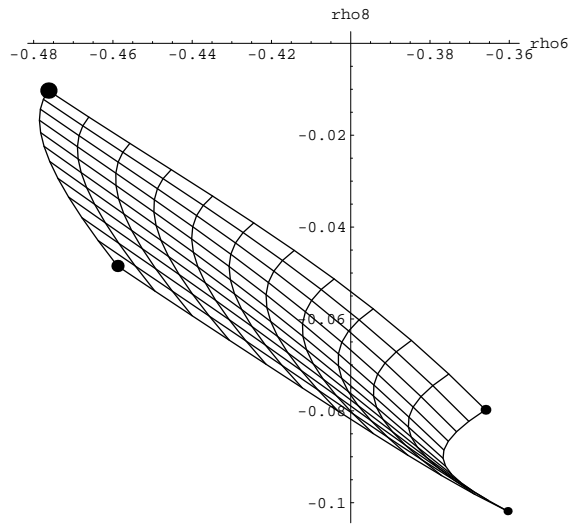}
     }
\caption{Same as Fig. 7, but projection of the image grid into the 
$\rho_6,\rho_8$ plane. Compare with Fig. 3.}
\end{figure}
   
\begin{figure}[htb] \centering
\mbox{  \epsfysize=100mm
         \epsffile{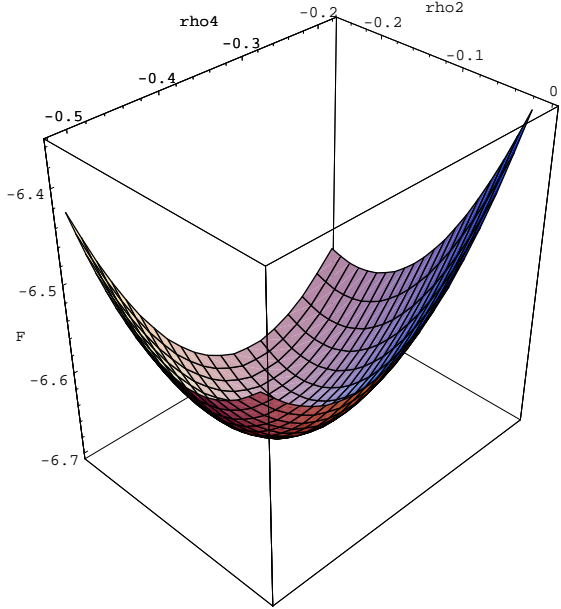}
     }
\caption{HK functional of the second toy model in $\{\rho_2,\rho_4\}$ frame.}
\end{figure}

Had we taken as the origin in density space the density of the HF solution 
for $u=0,$ different drifts of grid centers would have been observed. Such 
new drifts are likely to make better signals of true correlations in 
$\Psi.$ For the sake of comparison between the first and the second toy 
models we kept $\rho_0$ as a reference, but we have a direct access to the 
amount of true correlations: it is easy here to calculate the density 
{\it matrix} $\hat \rho,$ the diagonal of which gives $\rho.$ Here, for $u=0,$
the trace of $\hat \rho - \hat \rho^2$ is of order 5\%, a reasonable amount.
At that grid corner, $u_2=u_4=-2.2,$ the trace even reaches 8\%. It can be 
concluded that this second toy model does create correlations.

While $\rho_2$ and $\rho_4$ both vary by $\sim .3$ across their grid and 
$\rho_{10},\rho_{12}...$ can be neglected, the variation of $\rho_6$ and 
$\rho_8$ across the grid is of order $\sim .1,$ which is not so small. For 
the sake of comparison with Fig. 4 we now show in Figure 9 a plot of the HK 
functional in terms of $\{\rho_2,\rho_4\}$ again, but it will not be not 
forgotten in our agenda to find those two orthonormal combinations 
$\{ \sigma_2, \sigma_4 \}$ of $\rho_2,$ $\rho_4,$ $\rho_6$ and $\rho_8$ 
which allow the ``flattest'' projection of the grid. It is clear that, in 
the spirit of Eqs. (\ref{Fapp}) and (\ref{numFapp}), the best parametrization 
of $F$ should now be in terms of $\{ \sigma_2, \sigma_4 \}.$ In any case, 
from Fig. 9, the minimum of $F$ occurs at $\{\rho_2,\rho_4\}=\{-.14,-.36\},$ 
the grid center, as should be. Fig. 9 also shows deviations from a paraboloid 
clearly stronger than those of Fig. 4.

\section{Summary, discussion and conclusion}

Particle number conservation is essential to that key element in the proof of 
the HK theorem, the one-to-one correspondence between density and external 
potential. However, variations of the potential should not be trivial: they 
should differ from constants. We treated both constraints of i) matter 
conservation and ii) non triviality of potential variations on the same 
footing: a vanishing average for both the potential and the density 
variations. 

This constraint of vanishing average was implemented by means of a new family
of orthogonal polynomials; hence appeared a set of modes in both the density 
and the potential spaces. We proved, numerically with toy models, that such 
modes might have a physical meaning, on two counts, i) converse linear 
responses $\delta \rho/\delta u,$ $\delta u/\delta \rho$ might be reasonably 
simple when described in terms of such modes, and ii) the HK functional itself 
might be practically truncated into projections into subspaces spanned by a 
few modes.

We have not discussed in this paper the constraints of positivity of the 
density, but it is clear that, within an algebra of polynomials such as ours,
positivity conditions are not too difficult to implement. We have not 
discussed either more subtle constraints related to the Sobolev nature of
the topological spaces available for densities. For this question, we refer 
to \cite{Lieb}, \cite{Engli}, \cite{Leeuwen}. It can be stressed again that 
the compatibility of truncations of densities into a finite number of 
``polynomial modes'' with such fine constraints can easily be tested.

An ultimate goal would be to create a constructive theory of the HK 
functional. The functional differential equation, 
$\delta F/\delta \rho=-u[\rho],$ cannot be integrated in the density space
as long as $u[\rho]$ is not known accurately enough. Because of our detour
through many-body perturbation theories we are clearly far from 
the goal, but this work gives a frame in which the task should become easier.
The detour might allow a compression of the needed information through, for 
instance, the parametrization of a limited set of matrix elements of our 
matrices ${\cal D}$ for mean field approximations or $D$ in the correlated 
cases. Our main results are i) the existence of those special, orthogonal 
constrained polynomials $\Gamma_{\alpha}$ and associated modes $w_{\alpha}$ 
which design convenient sets of coordinates and convenient parametrizations
of $F,$ $\delta F/\delta \rho,$ $\delta^2 F/(\delta \rho\, \delta \rho'),...$ 
etc.,  ii) the explicit relation we showed between these modes and the 
traditional perturbation theories used in the many-body problem and iii) the 
likely possibility of truncated descriptions and accurate parametrizations.

Actually, in the context of extended systems, the idea of density waves
as important modes of the system has always been present. There remains to be 
seen, obviously, if our modes can be generalized to infinite systems. 
It also remains to be seen whether, for finite or infinite systems,
our truncations are always justified, whether infinite resummations are 
possible, whether collective degrees of freedom are present, or absent, 
because of our new representation. Also, because of our choice
of a Gaussian weight for the new family $\{\Gamma\},$ the present results
are better meaningful if restricted to nuclei. A generalization
to atoms and molecules obviously demands other weights for the constrained
polynomials. 

The usual perturbation theories have their hierarchy of modes in many-body 
space, most often a hierarchy of particle-hole components. 
Our approach, tuned to the one-body nature of the density functional,
replaces the particle-holes by other modes, in a transparent way.

It is a pleasure to thank Y. Abe, R. Balian and B. Eynard for stimulating 
discussions.

\end{document}